\documentclass[twocolumn,floatfix]{revtex4}

\usepackage{graphicx,amssymb}
\usepackage{hyperref}
\usepackage{amsmath}
\usepackage{setspace}

\begin{document}

\title{Emission of Terahertz Radiation from SiC}

\author{Jared H. Strait, Paul A. George, Jahan Dawlaty, Shriram Shivaraman, Mvs Chandrashekhar, Farhan Rana, and Michael G. Spencer}
\affiliation{School of Electrical and Computer Engineering, Cornell University, Ithaca, NY, 14853}

\begin{abstract}
We report the emission of strong coherent broadband terahertz radiation from 6H-Silicon-Carbide (SiC) excited with optical pulses.  The measured terahertz spectral signal-to-noise ratio is better than one thousand.  We determine that the terahertz radiation is generated via second order optical nonlinearity (optical rectification). We present a measurement of the ratio of nonlinear susceptibility tensor elements $\chi_{zzz}^{(2)}/\chi_{zxx}^{(2)}$ and the complex index of refraction of silicon carbide at THz frequencies.
\end{abstract}

\maketitle

Silicon Carbide (SiC) is a wide bandgap semiconductor possessing high mechanical stability, chemical stability, and thermal conductivity.  As a result, it is a promising candidate for high-field and high-power electronics \cite{sicbook,Yakimova06}, including applications intended for high temperature environments. Recently, SiC has also been explored for terahertz (THz) applications. Electrically-pumped THz emitters based on electronic transitions between impurity states have been demonstrated \cite{Powell05}. SiC devices, such as IMPATT oscillators, are also being explored for high power applications in the low THz region \cite{impatt}.

In this paper, we present results on the emission of coherent terahertz (THz) radiation from semi-insulating 6H-SiC excited with near-IR femtosecond optical pulses.  Broadband THz generation and detection in semiconductors with femtosecond (fs) optical pulses is a powerful and well-studied mechanism with applications in spectroscopy, imaging, and sensing \cite{Sakai}.  Many semiconductors, such as GaAs and InAs, emit coherent broadband THz pulses upon excitation with femtosecond optical pulses due to free-carrier generation and subsequent carrier dynamics in internal or externally applied electric fields \cite{Sakai,Gu02}. 6H-SiC has a large spontaneous polarization and, therefore, a large permanent bulk electric field \cite{sicpol}. However, the large indirect bandgap of SiC ($>3$ eV \cite{Lundquist95,Powers08}) implies that free-carrier generation via direct interband absorption is not possible. Free-carrier generation through two-photon or defect absorption is possible.  A nonlinear mechanism, such as optical rectification \cite{Sakai}, can also be responsible for the generation of THz radiation. The nonlinear optical properties of various SiC polytypes have been previously studied, and 6H-SiC is known to have a large second-order nonlinear susceptibility comparable to crystals such as lithium niobate and KTP \cite{Lundquist95, Bechstedt99}. In this paper, we study the THz radiation dependence on the optical pump polarization, the pump angle of incidence, and the pump power.  We show that second order optical nonlinearity,  and not free-carrier dynamics, is responsible for THz emission. Given its material hardness, high optical damage threshold, small optical losses, and high optical nonlinearity, SiC is promising for generating broadband high power THz radiation. 

\begin{figure}[tp]
  \centering
    \includegraphics[width=.44\textwidth]{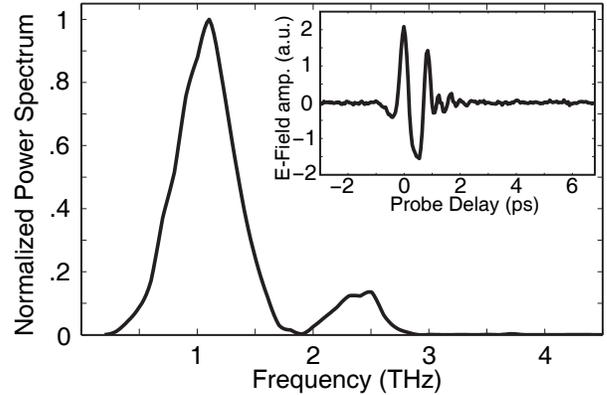}
  \caption{Power spectrum of observed electric field.  Inset: Observed transient electric field amplitude.}
  \label{fig:THz}
\end{figure}

For our experiments, we used a mode-locked Ti:Sapphire laser system to produce optical pump pulses with center wavelength $\sim$780 nm, duration $\sim$90 fs, and repetition rate 81 MHz.  Pulse energies varying from 1-15~nJ were used for excitation.  Pump polarization angle ($\theta$) was controlled with polarizers and a half-waveplate.  Vanadium-compensated semi-insulating 6H-SiC (0001) dies, 380 $\mu$m thick, were mounted vertically such that the crystals could be rotated about the vertical axis, thus controlling the angle of incidence ($\phi$) of the pump beam.  The pump beam was chopped at 2.5 kHz and focused onto the sample with a 25 mm focal length lens.  The emitted THz radiation was collected in a nitrogen-purged environment with off-axis parabolic mirrors and detected in the time domain by means of a standard electro-optic detection setup using a 1 mm thick (110)-ZnTe crystal \cite{Sakai,Zhang01}. Strong THz pulses with maximum spectral power signal-to-noise ratios better than one thousand and detection-limited bandwidths wider than 3 THz were observed.  A representative electric field transient and its accompanying spectrum are shown in Figure~\ref{fig:THz}.  The dip in the spectrum around 1.9 THz is discussed later. 

Terahertz emission via the generation and subsequent motion of free-carriers generally has little dependence on the pump polarization or the pump angle of incidence (other than that which is related to the transmission/reflection of the pump and the emitted THz radiation at the crystal interfaces). In contrast, the emission of THz radiation from nonlinear optical rectification is strongly dependent on the pump polarization and angle of incidence as dictated by the form of the second-order nonlinear susceptibility tensor $\chi_{ijk}^{(2)}$ \cite{Sakai,Boyd}. 6H-SiC has 6mm hexagonal crystal symmetry and therefore, under Kleinman's condition, there are only two independent components of $\chi_{ijk}^{(2)}$, written \cite{Boyd}
\begin{equation}
\chi_{ijk}^{(2)} = \left[ \begin{array}{cccccc}
0 & 0 & 0 & 0 & \chi_{zxx}^{(2)} & \;0\; \\
0 & 0 & 0 &  \chi_{zxx}^{(2)} & 0 & \;0\; \\
 \chi_{zxx}^{(2)} &  \chi_{zxx}^{(2)} &  \chi_{zzz}^{(2)} & 0 & 0 & \;0\;
\end{array} \right].
\label{eq:chi2}
\end{equation}

\begin{figure}[tb]
  \centering
    \includegraphics[width=.45\textwidth]{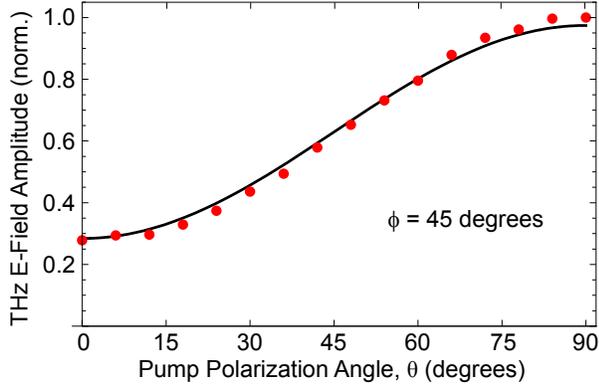}
  \caption{Normalized peak-to-peak electric field versus excitation polarization. 0 degrees corresponds to TE polarization, 90 degrees to TM.  Fit with $\chi_{zzz}^{(2)}/\chi_{zxx}^{(2)} = -3.0$.}
  \label{fig:Polarization}
\end{figure}
Figure~\ref{fig:Polarization} shows experimental results for the dependence of the THz E-field amplitude on the pump polarization while the pump angle of incidence is 45 degrees. The corresponding theoretical curve is also shown, calculated following the analysis of Chen \emph{et al.}\ \cite{Zhang01}.
Figure~\ref{fig:Polarization} shows a strong dependence on the pump polarization and a good fit of the optical rectification theory to the experimental data.  The ratio of $\chi_{zzz}^{(2)}/\chi_{zxx}^{(2)}$ is the only fitting parameter.  A value of $-3.0$ was used to obtain the fit in Figure~\ref{fig:Polarization}.

\begin{figure}[tb]
  \centering
    \includegraphics[width=.47\textwidth]{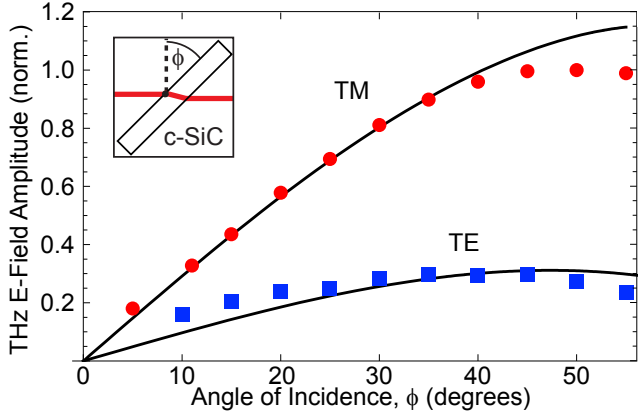}
  \caption{Normalized peak-to-peak electric field versus angle of incidence for TE (filled square) and TM (filled circle) pump polarizations.}
  \label{fig:Angle}
\end{figure}

To further confirm the optical rectification process, we study the THz electric field amplitude dependence on the pump angle of incidence.  The experimental results for TE and TM pump polarizations are shown in Figure~\ref{fig:Angle} along with the theoretical curves.  The theory is seen to fit the data well. As expected from the form of $\chi_{ijk}^{(2)}$, $E_{\rm THz}$ approaches zero at small angles of incidence.  Discrepancies at extreme angles of incidence are attributed to non-ideal phase matching and reduced collection efficiencies (to be discussed later).

\begin{figure}[tb]
  \centering
    \includegraphics[width=.45\textwidth]{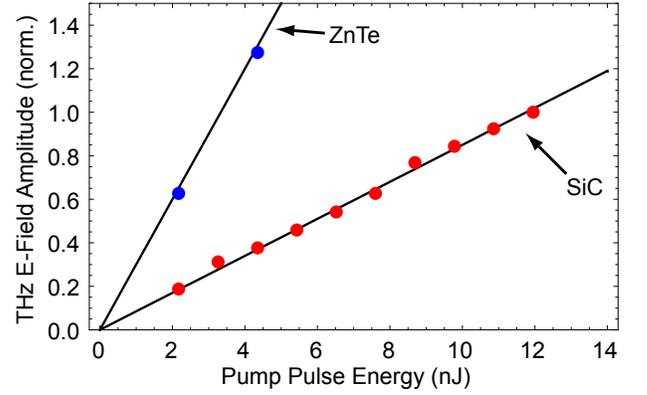}
  \caption{Normalized peak-to-peak electric field versus excitation pulse energy (at 780 nm) for SiC ($\phi=45^\circ$) and ZnTe (normal incidence).}
  \label{fig:Power}
\end{figure}

For a second order nonlinear emission process, the terahertz field amplitude is expected to be proportional to the pump power (or pump energy).  In Figure~\ref{fig:Power}, we plot the maximum THz pulse amplitude versus the pump pulse energy.  The observed linear dependence is in agreement with THz generation via optical rectification.  For comparison, we plot the amplitude of THz emission from a 1 mm thick (110) ZnTe crystal at normal incidence, adjusted for crystal length.

The results presented here indicate that the mechanism responsible for THz emission in SiC is nonlinear optical rectification.  The ratio of $\chi_{zzz}^{(2)}/\chi_{zxx}^{(2)}$, the fitting parameter used in Figures~\ref{fig:Polarization} and \ref{fig:Angle}, has been the subject of several theoretical and experimental investigations (see Table.I). The theoretically predicted value of this ratio is exactly -2 for cubic 3C-SiC and varies between -0.5 and -2.0 for $n$H-SiC (with $2\le n \le \infty$).
To date, the measured values of this ratio have been found to be much larger than the theoretically predicted values. Our measurements of the TM/TE ratio at $\phi=40$ degrees yielded a value of $\chi_{zzz}^{(2)}/\chi_{zxx}^{(2)}$ equal to $-3.0$ with a  95\% confidence interval of $\pm2.6$.  Table.I shows that our measured value of $\chi_{zzz}^{(2)}/\chi_{zxx}^{(2)}$ is the closest to the theoretical values among reported experimental results for 6H-SiC.

\begin{table}[htb]
\begin{tabular}{|cr|c|}
\hline\hline
&& $\chi_{zzz}^{(2)}/\chi_{zxx}^{(2)}$, 6H-SiC \\
\hline
Theory&Wu \emph{et al.} \cite{Wu08} & $-1.89$ \\
&Adolph \emph{et al.} \cite{Adolph00} & $-1.85$ \\
&Rashkeev \emph{et al.} \cite{Rashkeev98} & $-1.84$ \\
&Chen \emph{et al.} \cite{Chen94} & $-1.84$ \\
\hline
Experiment&Lundquist \emph{et al.} \cite{Lundquist95} & $-10$ \\
&Niedermeier \emph{et al.} \cite{Bechstedt99} & $-6$ \\
&This work & $-3.0\pm2.6$ \\
\hline\hline
\end{tabular}
\label{chiValues}
\caption{Summary of calculated (Wu \emph{et al.} \cite{Wu08}, Adolph \emph{et al.} \cite{Adolph00}, Rashkeev \emph{et al.} \cite{Rashkeev98}, Chen \emph{et al.} \cite{Chen94}) and measured (Lundquist \emph{et al.} \cite{Lundquist95}, Niedermeier \emph{et al.} \cite{Bechstedt99}) values of the ratio $\chi_{zzz}^{(2)}/\chi_{zxx}^{(2)}$. The measured value from this work is in close agreement with theoretical values to date.}
\end{table}

\begin{figure}[tb]
  \centering
    \includegraphics[width=.45\textwidth]{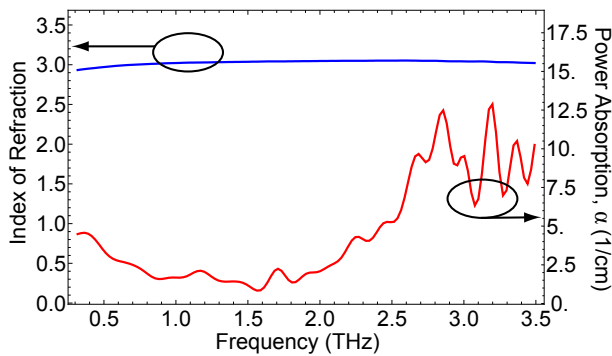}
  \caption{Index of refraction and power absorption of SiC versus frequency.}
  \label{fig:index}
\end{figure}

To study the efficacy of THz emission, we present measurements of the complex index of refraction of SiC at THz frequencies (Figure~\ref{fig:index}).  These measurements were obtained by transmitting broadband THz pulses generated with a photoconductive antenna through our samples.  We see that SiC is nearly dispersionless in the low THz range, with an index of refraction $n=3.0$.  Power absorption, given by $\alpha$, is also small in the 0.5-3.0 THz frequency range.  Since the index of SiC at $\sim$780 nm is $n\approx2.65$ \cite{Shaffer71}, there is non-ideal phase matching between the pump pulse and THz waves. The non-ideal phase matching could explain the discrepancy between theory and experiment at extreme angles in Figure~\ref{fig:Angle}.  At large $\phi$ angles, total internal reflection of the emitted THz radiation cone becomes significant.  Furthermore, since the THz radiation is not strictly collimated with the pump beam, a small amount of THz radiation is detectable even at angles near zero. The spectrum for difference frequency generation under non-ideal phase matching conditions is known to depend on the frequency \cite{Boyd}. Specifically, $E_{\rm THz}(\omega) \propto \sin(\Delta k L/2)/(\Delta k L/2)$, where $\Delta k = \omega(n_{\rm THz}-n_{\rm opt})/c$. According to this relation, the first zero in the spectrum should occur around 2.0 THz.  Figure~\ref{fig:THz} shows that the measured THz spectrum is in good agreement with the predictions.

In conclusion, we have demonstrated broadband coherent THz emission from SiC by optical rectification. THz emission via optical rectification has been well studied in zinc blende crystals with cubic symmetries, such as GaAs and ZnTe \cite{Rice94, Zhang01}, as well as in crystals of trigonal (LiNbO$_3$ \cite{Kuhl05}) and $\bar{6}$m2 hexagonal (GaSe \cite{Sasaki04}) symmetries.  Among these alternatives, ZnTe stands out as being particularly well phase-matched for THz generation \cite{Kuhl05}. However, its surface is known to burn under strong optical excitation and it also suffers from two-photon absorption because of its small direct bandgap ($\sim$2.25 eV). GaSe has a high optical damage threshold \cite{Schunemann98}, but the crystal is soft and fragile and its direct bandgap is also small ($\sim$2.12 eV). Given SiC's higher optical damage threshold \cite{Powers08}, comparable second order nonlinear susceptibility, robust mechanical properties, and large direct bandgap ($>$5 eV for 6H-SiC), it could prove to be a useful source of broadband THz radiation. Phase matching with a $\sim$780 nm pump pulse can be achieved with a tilted phase front of $28^\circ$ for efficient THz generation \cite{Kuhl05}.  

The authors acknowledge support from the National Science Foundation, the Air Force office of Scientific research contract No. FA9550-07-1-0332 (contract monitor Donald Silversmith) and Cornell Material Science and Engineering Center (CCMR) program of the National Science Foundation (cooperative agreement 0520404).

%
%
%
%
%

\end{document}